\documentclass[10pt]{article}
\usepackage{mathptmx}
\usepackage[T1]{fontenc}
\usepackage[latin9]{inputenc}
\usepackage[a4paper]{geometry}
\usepackage{amsmath}
\usepackage{amssymb}
\usepackage{esint}

\makeatletter

\usepackage{cite}

\usepackage{enumitem}\usepackage{subfigure}\usepackage{bbm}\usepackage{color}

\makeatother

\begin{document}

\title{Cosmological Power Spectrum in Noncommutative Space-time}

\author{Rahul Kothari, Pranati K. Rath and Pankaj Jain}

\maketitle
\begin{center}
{Dept. of Physics, Indian Institute of Technology Kanpur, Kanpur
- 208016, India} 
\par\end{center}

\begin{center}
\medskip{}
 Email: pranati@iitk.ac.in, rahulko@iitk.ac.in, pkjain@iitk.ac.in 
\par\end{center}
\begin{abstract}
We propose a generalized star product which deviates from the standard 
product when the fields 
at evaluated at different space-time points. This produces no changes in the
standard Lagrangian density in noncommutative space-time but produces
a change in the cosmological power spectrum. We show that the generalized
star product leads to physically consistent results and can fit
the observed data on hemispherical anisotropy in the cosmic microwave 
background radiation. 
\end{abstract}

\section{Introduction}

A remarkable prediction of quantum gravity is that space-time may
be noncommutative. The basic idea is that in order to probe short
distances we require higher energies. However at sufficiently high
energy we shall necessarily form black holes and hence loose precision
about space-time coordinates. This idea imposes some uncertainty relationships
among different coordinates which can be derived by proposing that
these coordinates do not commute \cite{Doplicher1994,Ahluwalia93,Connes1994,Madore1999,Landi1997,Bondia2001}.
It has been argued that this noncommutativity of coordinates might
have interesting implications for cosmology \cite{Greene2001,Lizzi2002,Brandenberger2002,Huang2003,Brandenberger2003,Bal2008,Barosi2008,Fatollahi2006,Fatollahi2006a,Akofor2008,Akofor2009}.
In particular the power spectrum generated during inflation could
be modified and may lead to signatures of non-Gaussianity 
\cite{Akofor2008,Akofor2009}. 

The non-commutative model is rather interesting since it has the potential
\cite{Koivisto11,Groeneboom2010,Rath2015} to explain the observed
hemispherical anisotropy in CMBR \cite{Eriksen2004,Eriksen2007,Hansen2009,Hoftuft2009,Erickcek2008,Planck2014a,Paci2013,Schmidt2013,Akrami2014}.
The reason is that it produces a dipolar term in the primordial power
spectrum. Such a term cannot arise with the framework of a standard
anisotropic model if we assume homogeneity. Assuming that it is possible
to generate the right form of the dipolar power spectrum starting
from a non-commutative model, which leads to physically acceptable
results, the consequences are mind boggling. It literally implies
that the shortest distance, perhaps Planck scale physics, associated
with the noncommutativity of space-time, may currently be probed at
the largest distance scales in the Universe. Furthermore anisotropies
(or inhomogeneities) at very early times may be observable today as
anisotropies on the largest distance scales \cite{Aluri2012,Pranati2013a}
and might be responsible for some of the observed anisotropies in
the Universe \cite{Jain1999,Hutsemekers1998,Costa2004,Ralston2004,Schwarz2004,Singal2011,Tiwari2013}
besides the hemispherical anisotropy \cite{Eriksen2004,Eriksen2007,Hansen2009,Hoftuft2009,Erickcek2008,Planck2014a,Paci2013,Schmidt2013,Akrami2014}.

The hemispherical anisotropy is parametrized in terms of the phenomenological
dipole modulation model \cite{Gordon2005,Gordon2007,Prunet2005,Bennett2011,Ghosh2016}.
It has been argued \cite{Koivisto11,Groeneboom2010,Rath2015} that
the power spectrum obtained in \cite{Akofor2009} is not
acceptable since it produces imaginary correlations among temperature
spherical harmonic coefficients, $a_{lm}$'s, while they should be
real. Clearly there is something wrong with the power
spectrum obtained in \cite{Akofor2009}. Some solutions to this
problem have already been proposed in Refs. \cite{Koivisto11,Rath2015}. 
However these do not really solve the problem. In particular 
the prescription given in \cite{Koivisto11} requires us to 
define the expectation value of different
parts of an operator differently. It is not clear how such a prescription
might emerge from a fundamental framework.  
Ref. \cite{Rath2015} instead suggests that we should take a different
product while computing the power spectrum. While this is permissible, 
it is ad hoc. It provides no theoretical justification for why a different 
product is used in the calculation of the power spectrum.

In the present paper we examine some of the assumptions that go into the
calculation of the power spectrum and subsequently the temperature
correlations. In their calculation the authors \cite{Akofor2008,Akofor2009}
assume that the transfer function which relates the power spectrum
in the early Universe is approximately the same as that assumed in
commutative space-times. This is reasonable since by the end of inflation
all effects of non-commutativity are expected to be negligible. Hence
the evolution can be well approximated by neglecting the effects of
non-commutativity.

The power spectrum in \cite{Akofor2008,Akofor2009} is obtained by
assuming that all products in noncommutative space-time must be taken
to be star products. This is also a reasonable assumption since a star
product implements the basic commutation relation among different
coordinates, given by \cite{Doplicher1994,Connes1994,Madore1999,Landi1997,Bondia2001},
\begin{equation}
[\hat{x}_{\mu},\hat{x}_{\nu}]=i\Theta_{\mu\nu}\,.\label{eq:noncom}
\end{equation}
Here the parameter, $\Theta_{\mu\nu}$ is antisymmetric and the coordinate
functions, $\hat{x}_{\mu}(x)$, depend on the choice of coordinate
system. Different choices will lead to different models of noncommutative
space-time. The authors \cite{Akofor2008,Akofor2009} consider a scalar
field theory in a background expanding Universe. The coordinates $\hat{x}_{\mu}$
are taken to be the comoving coordinates. They compute the two point
correlations of the scalar field, $\phi$, by assuming that their
product can be taken to be the star product.

We next point out that imposing the commutation relations on comoving
coordinates is simply a model. One can consider generalizations of
this model. Furthermore the commutation relations, Eq. \ref{eq:noncom},
provide guidance about the nature of the product rule only at leading
order in $\Theta_{\mu\nu}$. One could in principle have different
rules which differ from the star product at higher orders in $\Theta_{\mu\nu}$.
In our analysis, however, we shall be interested only in the leading
order term in $\Theta_{\mu\nu}$. At this order the product rule is
uniquely fixed by the commutation relations Eq. \ref{eq:noncom}.
Finally Eq. \ref{eq:noncom} provides guidance for the product rule
only when the two coordinates are same. In our analysis we require
the correlation function, 
\begin{equation}
\Delta(\vec{x},\vec{x}^{\,\prime})=\left\langle 0\left|\phi(\vec{x},t)
\star\phi(\vec{x}^{\,\prime},t)\right|0\right\rangle \,.\label{eq:correlation}
\end{equation}
i.e. the product of fields at two different spatial positions. For
such products we can examine a generalized product which involves
a form factor $F(x^{\mu}-x^{'\mu})$. Let us define a generalized star
product as 
\begin{equation}
\phi(\vec{x},t)\star\phi(\vec{x}^{\,\prime},t)=\exp\left(\frac{i}{2}F[(\vec{x}-
\vec{x}^{\,\prime})]\Theta^{\mu\nu}\frac{\overrightarrow{\partial}}{\partial x^{\mu}}\frac{\overrightarrow{\partial}}{\partial x'^{\nu}}\right)\phi(\vec{x},t)\phi(\vec{x}^{\,\prime},t)\label{eq:starproduct}
\end{equation}
Here we specialize to the case relevant to us, i.e. product of fields
at different spatial positions at same time. The definition has a
direct generalization for two different space-time positions. We need
to impose the constraint that in the limit 
$\vec{x}^{\,\prime}\rightarrow\vec{x}$,
$F[(\vec{x}-\vec{x}^{\,\prime})]\rightarrow1$, such that the
generalized star product reduces to the standard star product in this
limit. The exponential function is defined by its expansion. We may
make the rule that in each term in the expansion the form factor appears
on the left of the derivatives and hence does not get differentiated.
However as we shall see explicitly this is not necessary for the forms
we may choose for the form factor. All the derivative terms of the 
form factor cancel out in Eq. \ref{eq:starproduct}.

The proposed generalized star product is purely phenomenological. 
The form factor introduced will be chosen in order to fit the 
cosmological data. However it is theoretically well motivated since
such a form can, in principle, emerge from a fundamental framework. There is no
reason why for different space-time points the product must
be same as the standard star product. Furthermore,  
here we shall be interested only in the leading order contribution
in $\Theta^{\mu\nu}$. 
 Hence we expand the generalized star product
and keep only the leading order term. We obtain 
\begin{equation}
\phi(\vec{x},t)\star\phi(\vec{x}^{\,\prime},t)=\left(1+\frac{i}{2}F[(\vec{x}-
\vec{x}^{\,\prime})]\Theta^{\mu\nu}\frac{\overrightarrow{\partial}}{\partial x^{\mu}}\frac{\overrightarrow{\partial}}{\partial x'^{\nu}}\right)\phi(\vec{x},t)\phi(\vec{x}^{\,\prime},t)\,.
\label{eq:starproduct1}
\end{equation}
We clarify that at higher orders the form of the product may deviate 
from the exponential form proposed in Eq. \ref{eq:starproduct}. 
Indeed our fit to the cosmological data probes this product only at this
order. Hence we can trust our proposed form only at first order in
the parameter $\theta^{\mu\nu}$.  
In the next section we compute the power spectrum of the scalar field
making a suitable choice of the form factor $F[(\vec{x}-\vec{x}^{\,\prime})]$.

\subsection{Power spectrum in FRW background}

In this section we compute the correlation function $\Delta(\vec{x},\vec{x}^{\,\prime})$
defined in Eq. \ref{eq:correlation} for the case of an expanding
de Sitter Universe at leading order in $\theta^{\mu\nu}$. The scalar
field may be expressed as, 
\begin{equation}
\phi\left(\vec{x},t\right)=\int\frac{d^{3}\vec{k}}{\left(2\pi\right)^{3}}
\left(a_{\vec{k}}e^{i\vec{k}\cdot\vec{x}}\zeta_{\vec{k}}\left(t\right)+a_{\vec{k}}^{\dagger}e^{-i\vec{k}\cdot\vec{x}}\zeta_{\vec{k}}^{\star}\left(t\right)\right)
\end{equation}
where $\zeta_{\vec{k}}=u_{\vec{k}}/a$ and the mode function
$u_{\vec{k}}=\frac{e^{-ik\eta}}{\sqrt{2k}}\left(1-\frac{i}{k\eta}\right)$.
A direct calculation yields 
\begin{equation}
\Delta(\vec{x}-\vec{x}^{\,\prime})=\left\langle 0\left|\phi\left(\vec{x},t\right)\phi\left(\vec{x}^{\,\prime},t\right)\right|0\right\rangle +\Delta_{1}(\vec{x}-
\vec{x}^{\,\prime})\,.
\end{equation}
where we have used translational invariance and set $\Delta(\vec{x},\vec{x}^{\,\prime})=\Delta(\vec{x}-\vec{x}^{\,\prime})$.
Here the first term on the right hand side is the standard contribution
in commutative space-time and the second term is the leading order
correction. We are interested in its Fourier transform, 
\begin{equation}
\delta P(\vec{k})=\int d^{3}\vec{X}e^{-i\vec{k}\cdot\vec{X}}\Delta_{1}(\vec{x}-
\vec{x}^{\,\prime}),
\end{equation}
where $\vec{X}=\vec{x}-\vec{x}^{\,\prime}$. We obtain 
\begin{equation}
\delta P(\vec{k})=\frac{1}{2}\Theta^{0i}\int\frac{d^{3}\vec{X}d^{3}\vec{q}}{(2\pi)^{3}}e^{i(\vec{q}-\vec{k})\cdot\vec{X}}F(\vec{X})q_{i}f(q),
\end{equation}
where 
\begin{equation}
f(q)=\dot{\zeta}_{\vec{q}}\zeta_{\vec{q}}^{\star}+\dot{\zeta}_{\vec{q}}^{\star}\zeta_{\vec{q}}=-\frac{2H^{3}}{q^{3}}
\end{equation}
and $H$ is the Hubble's constant.

We next make the following choice for the form factor: 
\begin{equation}
F(\vec{X})=\cos(\vec{\lambda}\cdot\vec{X}/\eta)+iB\vec{X}\cdot\vec{X}
\end{equation}
where $B$ and $\lambda$ are parameters. 
The form factor goes to one in the limit $\vec{X}\rightarrow0$.
It is also clear that 
\begin{equation}
\Theta^{ij}\frac{\partial}{\partial x^{i}}\frac{\partial}{\partial x^{'j}}F(\vec{X})=0
\end{equation}
This is because $F(\vec{X})$ is an even function of $\vec{X}$.
Hence all the terms in the expansion of the exponential in Eq. \ref{eq:starproduct}
which involve derivatives of $F(\vec{X})$ vanish. After computation
one finds that 
\[
\delta P(\vec{k})=-\frac{\eta^{2}H^{3}\Theta^{0i}}{2}\left[\frac{\left(\eta k_{i}-\lambda_{i}\right)}{\left|\eta\vec{k}-\vec{\lambda}\right|^{3}}+\frac{\left(\eta k_{i}+\lambda_{i}\right)}{\left|\eta\vec{k}+\vec{\lambda}\right|^{3}}\right]+i\frac{6iBH^{3}\Theta^{0i}k_{i}}{k^{5}}
\]
In the limit $\eta\rightarrow0$ we obtain 
\begin{equation}
\delta P(\vec{k})=i\frac{6iBH^{3}\Theta^{0i}k_{i}}{k^{5}}
\end{equation}
Hence we obtain a power spectrum of the form which was anticipated
in \cite{Koivisto11,Rath2015}. The imaginary part of the form factor
has been chosen so that we obtain the power required to fit the data
\cite{Kothari2016,Ghosh2016}. We point out that the  
dipolar power spectrum was found to decay by approximately one power
of $k$ higher than the standard scale invariant power spectrum 
\cite{Kothari2016}. This
is exactly what is found in our analysis. 
Our analysis shows that it is possible for a non-commutative model
to produce a power spectrum required for a satisfactory fit to the
hemispherical anisotropy in a homogeneous Universe. Our model is
phenomenological and it remains to be determined whether such a
model can arise from a fundamental framework. 

\section{Conclusion}

We have proposed a generalized star product which is applicable when
the fields are evaluated at two different space-time positions, $x_1^\mu$
and $x_2^\mu$. 
The product involves an effective form factor which becomes equal 
to unity when $x_1^\mu = x_2^\mu$. Using a model for this 
form factor we compute the cosmological primordial power spectrum 
produced during inflation. We find that in Fourier space the power 
spectrum acquires a dipolar imaginary structure exactly as 
anticipated in \cite{Koivisto11,Rath2015}.
Such a structure is required in order that it yield an acceptable
CMB temperature anisotropy pattern. It has been argued that this might
provide an explanation of the observed hemispherical anisotropy \cite{Eriksen2004,Eriksen2007,Hansen2009,Hoftuft2009,Erickcek2008,Planck2014a,Paci2013,Schmidt2013,Akrami2014}
or equivalently the dipole modulation \cite{Gordon2005,Gordon2007,Prunet2005,Bennett2011,Pranati2013b,Rath2014}
of CMB temperature. 
Our results show that it is possible to explain the hemispherical
anisotropy in terms of noncommutative space-time.

\bigskip{}
 \textbf{Acknowledgment}: Rahul Kothari sincerely acknowledges CSIR, 
New Delhi for
the award of fellowship during the work.

\end{document}